\documentclass[twoside]{article}%
\usepackage{amssymb}
\usepackage{amsfonts}
\usepackage{amsmath}
\usepackage{graphicx}%
\providecommand{\U}[1]{\protect\rule{.1in}{.1in}}
\newtheorem{theorem}{Theorem}

\newtheorem{corollary}[theorem]{Corollary}

\newtheorem{lemma}[theorem]{Lemma}

\newenvironment{proof}[1][Proof]{\noindent\textbf{#1.} }{\ \rule{0.5em}{0.5em}}
\begin{document}

\title{\textbf{Analytical Solutions to the Navier-Stokes Equations with
Density-dependent Viscosity and with Pressure}}
\author{\textsc{Ling Hei Yeung\thanks{E-mail address: lightisgood2005@yahoo.com.hk}}\\\textit{Department of Mathematics, The Hong Kong Baptist University,}\\\textit{Kowloon Tong, Hong Kong}
\and Y\textsc{uen} M\textsc{anwai\thanks{E-mail address: nevetsyuen@hotmail.com }}\\\textit{Department of Applied Mathematics, The Hong Kong Polytechnic
University,}\\\textit{Hung Hom, Kowloon, Hong Kong}}
\date{Revised 20-Feb-2009}
\maketitle

\begin{abstract}
This article is the continued version of the analytical solutions for the
pressureless Navier-Stokes equations with density-dependent viscosity
\cite{Y2}. We are able to extend the similar solutions structure to the case
with pressure under some restriction to the constants $\gamma$ and $\theta$.


\end{abstract}

\section{Introduction}

The Navier-Stokes equations can be formulated in the following form:%
\begin{equation}
\left\{
\begin{array}
[c]{rl}%
{\normalsize \rho}_{t}{\normalsize +\nabla\cdot(\rho u)} & {\normalsize =}%
{\normalsize 0,}\\
{\normalsize (\rho u)}_{t}{\normalsize +\nabla\cdot(\rho u\otimes
u)+\delta\nabla P} & {\normalsize =}vis(\rho,u).
\end{array}
\right.  \label{eq1}%
\end{equation}
As usual, $\rho=\rho(x,t)$ and $u(x,t)$ are the density, the velocity
respectively. $P=P(\rho)$ is the pressure. We use a $\gamma$-law on the
pressure, i.e.
\begin{equation}
P(\rho)=K\rho^{\gamma}, \label{eq2}%
\end{equation}
with $K>0$, which is a universal hypothesis. The constant $\gamma=c_{P}%
/c_{v}\geq1$, where $c_{p}$ and $c_{v}$ are the specific heats per unit mass
under constant pressure and constant volume respectively, is the ratio of the
specific heats. $\gamma$ is the adiabatic exponent in (\ref{eq2}). In
particular, the fluid is called isothermal if $\gamma=1$. It can be used for
constructing models with non-degenerate isothermal fluid. $\delta$ can be the
constant $0$ or $1$. When $\delta=0$, we call the system is pressureless; when
$\delta=1$, we call that it is with pressure. And $vis(\rho,u)$ is the
viscosity function. When $vis(\rho,u)=0$, the system (\ref{eq1}) becomes the
Euler equations. For the detailed study of the Euler and Navier-Stokes
equations, see \cite{CW} and \cite{L}. Here we consider the density-dependent
viscosity function as follows:%
\[
vis(\rho,u)\doteq\bigtriangledown(\mu(\rho)\bigtriangledown\cdot u)\text{.}%
\]
where $\mu(\rho)$ is a density-dependent viscosity function, which is usually
written as $\mu(\rho)\doteq\kappa\rho^{\theta}$ with the constants $\kappa,$
$\theta>0$. For the study of this kind of the above system, the readers may
refer \cite{Ni}, \cite{Y1}.

The Navier-Stokes equations with density-dependent viscosity in radial
symmetry can be expressed by:%
\begin{equation}
\left\{
\begin{array}
[c]{rl}%
\rho_{t}+u\rho_{r}+\rho u_{r}+{\normalsize \frac{N-1}{r}\rho u} &
{\normalsize =0,}\\
\rho\left(  u_{t}+uu_{r}\right)  +\nabla K\rho^{\gamma} & {\normalsize =(}%
\kappa\rho^{\theta})_{r}\left(  \frac{N-1}{r}u+u_{r}\right)  +\kappa
\rho^{\theta}(u_{rr}+\frac{N-1}{r}u_{r}-\frac{N-1}{r^{2}}u),
\end{array}
\right.  \label{gOAL}%
\end{equation}

Recently, Yuen's results \cite{Y2} showed that there exists a family of the
analytical solutions for the pressureless Navier-Stokes equations with
density-dependent viscosity:\newline for $\theta=1$:%
\begin{equation}
\left\{
\begin{array}
[c]{c}%
\rho(t,r)=\frac{e^{y(r/a(t))}}{a(t)^{N}},\text{ }u(t,r)=\frac{\overset{\cdot
}{a}(t)}{a(t)}r,\\
\overset{\cdot\cdot}{a}(t)=\frac{\lambda\overset{\cdot}{a}(t)}{a(t)^{2}%
},a(0)=a_{0}>0,\overset{\cdot}{a}(0)=a_{1},\\
y(x)=\frac{\lambda}{2N\kappa}x^{2}+\alpha,
\end{array}
\right.
\end{equation}
for $\theta\neq1$:%
\begin{equation}
\left\{
\begin{array}
[c]{c}%
\rho(t,r)=\left\{
\begin{array}
[c]{cc}%
\frac{^{y(r/a(t))}}{a(t)^{N}}, & \text{ for }y(\frac{r}{a(t)})\geq0;\\
0, & \text{for }y(\frac{r}{a(t)})<0
\end{array}
\right.  ,\text{ }u(t,r)=\frac{\overset{\cdot}{a}(t)}{a(t)}r,\\
\overset{\cdot\cdot}{a}(t)=\frac{-\lambda\overset{\cdot}{a}(t)}%
{a(t)^{^{N\theta-N+2}}},\text{ }a(0)=a_{0}>0,\text{ }\overset{\cdot}%
{a}(0)=a_{1},\\
y(x)=\sqrt[\theta-1]{\frac{1}{2}(\theta-1)\frac{-\lambda}{N\kappa\theta}%
x^{2}+\alpha^{\theta-1}},
\end{array}
\right.
\end{equation}
where $\alpha>0$.

In this article, we extend the results form the study of the analytical
solutions in the Navier-Stokes equations without pressure \cite{Y2} to the
case with pressure. The techniques of separation method of self-similar
solutions were also found to treat other similar systems in \cite{DXY},
\cite{Li}, \cite{M1}, \cite{Y}, \cite{Y1} and \cite{Y2}.

Our main result is the following theorem:

\begin{theorem}
\label{thm:1}For the $N$-dimensional Navier-Stokes equations in radial
symmetry (\ref{gOAL}), there exists a family of solutions, those are:\newline
for $\theta=\gamma=1$,%
\begin{equation}
\left\{
\begin{array}
[c]{c}%
\rho(t,r)=\frac{Ae^{B(\frac{r}{a(t)})^{2}+C}}{a(t)^{N}},\text{ }%
u(t,r)=\frac{\overset{\cdot}{a}(t)}{a(t)}r,\\
\ddot{a}(t)-\frac{2BK}{a(t)}+\frac{BN\kappa\overset{\cdot}{a}(t)}{a(t)^{2}%
}=0,\text{ }a(0)=a_{0}>0,\text{ }\overset{\cdot}{a}(0)=a_{1},
\end{array}
\right.  \label{YeungYuen1}%
\end{equation}
where $A\geq0$, $B$ and $C$ are constants.\newline for $\theta=\gamma>1,$%
\begin{equation}
\left\{
\begin{array}
[c]{c}%
\rho(t,r)=\left\{
\begin{array}
[c]{c}%
\frac{y(\frac{r}{a(t)})}{a(t)^{N}}\text{, for }y(\frac{r}{a(t)})\geq0\\
0\text{, for }y(\frac{r}{a(t)})<0
\end{array}
\right.  ,\text{ }u(t,r)=\frac{\overset{\cdot}{a}(t)}{a(t)}r,\\
\frac{\overset{\cdot\cdot}{a}(t)}{a(t)^{N}}+\frac{K\gamma}{a(t)^{\theta N+1}%
}-\frac{N\kappa\theta\overset{\cdot}{a}(t)}{a(t)^{^{\theta N+2}}}=0,\text{
}a(0)=a_{0}>0,\text{ }\overset{\cdot}{a}(0)=a_{1},\\
y(z)=\sqrt[\theta-1]{\frac{1}{2}(\theta-1)z^{2}+\alpha^{\theta-1}},
\end{array}
\right.  \label{YY2}%
\end{equation}
where $a_{0}$, $a_{1}$ and $\alpha>0$ are constants$;$\newline for
$\frac{\gamma}{2}+\frac{1}{2}-\frac{1}{N}=\theta\geq1-\frac{1}{N}$,%
\begin{equation}
\left\{
\begin{array}
[c]{c}%
\begin{array}
[c]{c}%
\rho(t,r)=\left\{
\begin{array}
[c]{c}%
\frac{y(\frac{r}{a(t)})}{a(t)^{N}}\text{, for }y(\frac{r}{a(t)})\geq0\\
0\text{, for }y(\frac{r}{a(t)})<0
\end{array}
\right.  ,\text{ }u(t,r)=\frac{\overset{\cdot}{a}(t)}{a(t)}r\\
a(t)=\sigma(mt+n)^{s}\text{, }0<s=\frac{2}{\gamma N-N+2}\leq1
\end{array}
\\
\left[  \frac{K\gamma}{s\sigma^{\gamma N+1}}y(z)^{\gamma-2}-\frac
{mN\kappa\theta}{\sigma^{\theta N+1}}y(z)^{\theta-2}\right]  \dot{y}%
(z)=\frac{(1-s)m^{2}}{\sigma^{N-1}}z\text{, }y(0)=\alpha>0
\end{array}
\right.  \label{yy3}%
\end{equation}
where $m$, $n>0$, $\sigma>0$ and $\alpha$ are constants.
\end{theorem}

\section{Separation Method of Self-Similar Solutions}

Before we present the proof of Theorem \ref{thm:1}, Lemmas 3 and 12 of
\cite{Y2} are quoted here.

\begin{lemma}
[Lemma 3 of \cite{Y2}]\label{lem:generalsolutionformasseq}For the equation of
conservation of mass in radial symmetry:
\begin{equation}
\rho_{t}+u\rho_{r}+\rho u_{r}+\frac{N-1}{r}\rho u=0,
\label{massequationspherical}%
\end{equation}
there exist solutions,%
\begin{equation}
\rho(t,r)=\frac{f(r/a(t))}{a(t)^{N}},\text{ }{\normalsize u(t,r)=}%
\frac{\overset{\cdot}{a}(t)}{a(t)}{\normalsize r,}
\label{generalsolutionformassequation}%
\end{equation}
with the form $f\geq0\in C^{1}$ and $a(t)>0\in C^{1}.$
\end{lemma}

\begin{lemma}
[Lemma 12 of \cite{Y2}]$\label{lemma4455}$For the ordinary differential
equation%
\begin{equation}
\left\{
\begin{array}
[c]{c}%
\overset{\cdot}{y}(z)y(z)^{n}-\xi x=0,\\
y(0)=\alpha>0,n\neq-1,
\end{array}
\right.  \label{seperateODE}%
\end{equation}
where $\xi$ and $n$ are constants,\newline we have the solution%
\begin{equation}
y(z)=\sqrt[n+1]{\frac{1}{2}(n+1)\xi z^{2}+\alpha^{n+1}},
\end{equation}
where the constant $\alpha>0$.
\end{lemma}

At this stage, we can show the proof of Theorem \ref{thm:1}.

\begin{proof}
Our solutions (\ref{YeungYuen1}), (\ref{YY2}) and (\ref{yy3}) fit the mass
equation (\ref{gOAL})$_{1}$ by Lemma (\ref{lem:generalsolutionformasseq}).
Next, for the equation (\ref{gOAL})$_{2}$, we plug our solutions to check
that.\newline For $\theta=\gamma=1$, we get%
\begin{align}
&  \rho(u_{t}+u\cdot u_{r})+K\left[  \rho\right]  _{r}-{\normalsize (}%
\kappa\rho)_{r}\left(  \frac{N-1}{r}u+u_{r}\right)  -\kappa\rho_{r}%
(u_{rr}+\frac{N-1}{r}u_{r}-\frac{N-1}{r^{2}}u)\\
&  =\frac{Ae^{B(\frac{r}{a(t)})^{2}+C}}{a(t)^{3}}\frac{\overset{\cdot\cdot}%
{a}(t)}{a(t)}r+K\frac{Ae^{B(\frac{r}{a(t)})^{2}+C}}{a(t)^{4}}B\left[
\frac{-2r}{a(t)}\right]  -\frac{AN\kappa e^{B(\frac{r}{a(t)})^{2}+C}}%
{a(t)^{4}}B\left[  \frac{-2r}{a(t)}\right]  \frac{\overset{\cdot}{a}(t)}%
{a(t)}\\
&  =\frac{Ae^{B(\frac{r}{a(t)})^{2}+C}r}{a(t)^{4}}\left[  \ddot{a}%
(t)-\frac{2BK}{a(t)}+\frac{BN\kappa\overset{\cdot}{a}(t)}{a(t)^{2}}\right]  \\
&  =0,
\end{align}
where the function $a(t)$ is required to be
\begin{equation}
\ddot{a}(t)-\frac{2BK}{a(t)}+\frac{BN\kappa\overset{\cdot}{a}(t)}{a(t)^{2}%
}=0,\text{ }a(0)=a_{0}>0,\text{ }\overset{\cdot}{a}(0)=a_{1},
\end{equation}
where $a_{0}$ and $a_{1}$ are constants.

For $\theta=\gamma>1$, we have:
\begin{align}
&  \rho(u_{t}+u\cdot u_{r})+K\left[  \rho^{\gamma}\right]  _{r}%
-{\normalsize (}\kappa\rho^{\theta})_{r}\left(  \frac{N-1}{r}u+u_{r}\right)
-\kappa\rho_{r}^{\theta}(u_{rr}+\frac{N-1}{r}u_{r}-\frac{N-1}{r^{2}}u)\\
&  =\frac{y(\frac{r}{a(t)})}{a(t)^{N}}\frac{\overset{\cdot\cdot}{a}(t)}%
{a(t)}r+\frac{K\theta y(\frac{r}{a(t)})^{\theta-1}\dot{y}(\frac{r}{a(t)}%
)}{a(t)^{\gamma N+1}}-\frac{N\kappa\theta y(\frac{r}{a(t)})^{\theta-1}\dot
{y}(\frac{r}{a(t)})}{a(t)^{^{\theta N+2}}}\overset{\cdot}{a}(t).\label{eq1111}%
\end{align}
By defining $z:=r/a(t)$, and requiring
\begin{align}
y(z)z-y(z)^{\theta-1}\dot{y}(z) &  =0,\\
z-y(z)^{\theta-2}\dot{y}(z) &  =0,\label{eq2222}%
\end{align}
(\ref{eq1111}) becomes%
\begin{equation}
=y(z)z\left[  \frac{\overset{\cdot\cdot}{a}(t)}{a(t)^{N}}+\frac{K\theta
}{a(t)^{\theta N+1}}-\frac{N\kappa\theta\overset{\cdot}{a}(t)}{a(t)^{^{\theta
N+2}}}\right]  =0,
\end{equation}
where the function $a(t)$ is required to be
\begin{equation}
\frac{\overset{\cdot\cdot}{a}(t)}{a(t)^{N}}+\frac{K\gamma}{a(t)^{\theta N+1}%
}-\frac{N\kappa\theta\overset{\cdot}{a}(t)}{a(t)^{^{\theta N+2}}}=0.
\end{equation}
Therefore, the equation (\ref{gOAL})$_{2}$ is satisfied.\newline With
$n:=\theta-2$ and $\xi:=1$, in Lemma \ref{lemma4455}, (\ref{eq2222}) can be
solved by
\begin{equation}
y(z)=\sqrt[\theta-1]{\frac{1}{2}(\theta-1)z^{2}+\alpha^{\theta-1}},
\end{equation}
where $\alpha>0$ is a constant.\newline For the case of $\frac{\gamma}%
{2}+\frac{1}{2}-\frac{1}{N}=\theta\geq1-\frac{1}{N}$, we have,%
\begin{align}
&  \rho(u_{t}+u\cdot u_{r})+K\left[  \rho^{\gamma}\right]  _{r}%
-{\normalsize (}\kappa\rho^{\gamma})_{r}\left(  \frac{N-1}{r}u+u_{r}\right)
-\kappa\rho_{r}^{\theta}(u_{rr}+\frac{N-1}{r}u_{r}-\frac{N-1}{r^{2}}u)\\
&  =\frac{y(\frac{r}{a(t)})}{a(t)^{N}}\frac{\overset{\cdot\cdot}{a}(t)}%
{a(t)}r+\frac{K\theta y(\frac{r}{a(t)})^{\gamma-1}\dot{y}(\frac{r}{a(t)}%
)}{a(t)^{\gamma N+1}}-\frac{N\kappa\theta y(\frac{r}{a(t)})^{\theta-1}\dot
{y}(\frac{r}{a(t)})}{a(t)^{^{\theta N+2}}}\overset{\cdot}{a}(t).
\end{align}
By letting $a(t)=\sigma(mt+n)^{s}$, we have%
\begin{align}
&  =y(\frac{r}{a(t)})\frac{s(s-1)(mt+n)^{s-2}}{\sigma^{N}(mt+n)^{sN}}%
\frac{m^{2}\sigma r}{a(t)}+\frac{K\theta y(\frac{r}{a(t)})^{\gamma-1}\dot
{y}(\frac{r}{a(t)})}{\sigma^{\gamma N+1}(mt+n)^{s(\gamma N+1)}}-\frac{sm\sigma
N\kappa\theta y(\frac{r}{a(t)})^{\theta-1}\dot{y}(\frac{r}{a(t)})}%
{\sigma^{\theta N+2}(mt+n)^{s(\theta N+2)}}(mt+n)^{s-1}\\
&  =y(\frac{r}{a(t)})\frac{s(s-1)m^{2}r}{\sigma^{N-1}a(t)}\frac{1}%
{(mt+n)^{sN-(s-2)}}+\frac{K\theta y(\frac{r}{a(t)})^{\gamma-1}\dot{y}(\frac
{r}{a(t)})}{\sigma^{\gamma N+1}(mt+n)^{s(\gamma N+1)}}-\frac{smN\kappa\theta
y(\frac{r}{a(t)})^{\theta-1}\dot{y}(\frac{r}{a(t)})}{\sigma^{\theta
N+1}(mt+n)^{s(\theta N+2)-(s-1)}}\\
&  =y(\frac{r}{a(t)})\frac{s(s-1)m^{2}r}{\sigma^{N-1}a(t)}\frac{1}%
{(mt+n)^{sN-s+2}}+\frac{K\theta y(\frac{r}{a(t)})^{\gamma-1}\dot{y}(\frac
{r}{a(t)})}{\sigma^{\gamma N+1}(mt+n)^{s(\gamma N+1)}}-\frac{smN\kappa\theta
y(\frac{r}{a(t)})^{\theta-1}\dot{y}(\frac{r}{a(t)})}{\sigma^{\theta
N+1}(mt+n)^{\theta N+s+1}}.\label{ww11}%
\end{align}
Here we require that%
\begin{equation}
\left\{
\begin{array}
[c]{c}%
sN-s+2=s(\gamma N+1),\\
s(\gamma N+1)=s(\theta N+1)+1.
\end{array}
\right.
\end{equation}
That is
\begin{equation}
0<s=\frac{1}{(\gamma-\theta)N}=\frac{2}{\gamma N-N+2}\leq1.\label{rr1}%
\end{equation}
In the solutions (\ref{yy3}), it fits to the conditions (\ref{rr1}) by setting
$\frac{\gamma}{2}+\frac{1}{2}-\frac{1}{N}=\theta\geq1-\frac{1}{N}>0$ and
$s=\frac{2}{\gamma N-N+2}$. Additionally by defining $z:=r/a(t)$, the equation
(\ref{ww11}) becomes%
\begin{equation}
=\frac{y(z)s}{(mt+n)^{Ns-s+2}}\left[  \frac{(s-1)m^{2}}{\sigma^{N-1}}%
z+\frac{K\gamma}{s\sigma^{\gamma N+1}}y(z)^{\gamma-2}\dot{y}(z)-\frac
{mN\kappa\theta}{\sigma^{\theta N+1}}y(z)^{\theta-2}\dot{y}(z)\right]  ,
\end{equation}
Here we require that
\begin{equation}
\left[  \frac{K\gamma}{s\sigma^{\gamma N+1}}y(z)^{\gamma-2}-\frac
{mN\kappa\theta}{\sigma^{\theta N+1}}y(z)^{\theta-2}\right]  \dot{y}%
(z)=\frac{(1-s)m^{2}}{\sigma^{N-1}}z.\label{iii1}%
\end{equation}
The proof is completed.
\end{proof}

In the corollary can be followed immediately:

\begin{corollary}
For $m<0$, the solutions (\ref{yy3}) blowup at the finite time $T=-m/n$.
\end{corollary}

\end{document}